# HYDRODYNAMIC EVOLUTION OF COALESCING COMPACT BINARIES

FREDERIC A. RASIO
*Institute for Advanced Study*
*Princeton, NJ 08540, USA*

AND

STUART L. SHAPIRO
*Center for Radiophysics and Space Research*
*Cornell University, Ithaca, NY 14853, USA*

**Abstract.** In addition to their possible relevance to gamma-ray bursts, coalescing binary neutron stars have long been recognized as important sources of gravitational radiation that should become detectable with the new generation of laser interferometers such as LIGO. Hydrodynamics plays an essential role near the end of the coalescence when the two stars finally merge together into a single object. The shape of the corresponding burst of gravitational waves provides a direct probe into the interior structure of a neutron star and the nuclear equation of state. The interpretation of the gravitational waveform data will require detailed theoretical models of the complicated three-dimensional hydrodynamic processes involved. Here we review the results of our recent work on this problem, using both approximate quasi-analytic methods and large-scale numerical hydrodynamics calculations on supercomputers. We also discuss briefly the coalescence of white-dwarf binaries, which are also associated with a variety of interesting astrophysical phenomena.

## 1. Introduction

The coalescence and merging of two stars into a single object is the almost inevitable end-point of compact binary evolution. Dissipation mechanisms such as the emission of gravitational radiation are always present and cause the binary orbit to decay. The terminal stage of this orbital decay is always hydrodynamic in nature, with the final merging of the two stars taking place on a time scale comparable to the rotation period. In some systems, this is because the transfer of mass from one component to the other can become dynamically unstable, with the mass donor undergoing complete tidal disruption eventually. In addition, it was realized recently that, even if the mass transfer is stable or does not occur, *global hydrodynamic instabilities* can drive the binary system to rapid coalescence once the tidal interaction between the two stars becomes sufficiently strong (Rasio & Shapiro 1992, 1994, 1995a,b hereafter RS1–4; Lai, Rasio, & Shapiro 1993a,b, 1994a,b,c, hereafter LRS1–5 or collectively LRS). Using numerical hydrodynamic calculations, we demonstrated for the first time in RS1 the existence of these global instabilities for binary systems containing a compressible fluid. In addition, the classical *analytic* work for binaries containing an *incompressible* fluid (Chandrasekhar 1969) was extended to compressible fluids in the work of LRS. This new analytic study confirmed the existence of dynamical and secular instabilities for sufficiently close binary systems containing polytropes. However, numerical calculations remain essential for establishing the stability limits of close binaries accurately and for following the nonlinear evolution of unstable systems all the way to complete coalescence.

This review will concentrate on the coalescence of compact binaries, containing either two neutron stars (§2) or two white dwarfs (§3). Many of the results for white dwarfs, however, are also relevant to low-mass main-sequence stars in contact systems and the problem of blue straggler formation through binary coalescence (RS3)

## 2. Coalescing Neutron Star Binaries

### 2.1. ASTROPHYSICAL MOTIVATION

Coalescing neutron-star binaries are the most promising known sources of gravitational radiation that could be detected by the new generation of laser interferometers such as the Caltech-MIT LIGO (Thorne 1987; Abramovici et al. 1992; Cutler et al. 1992) and the European VIRGO (Bradaschia et al. 1990). Statistical arguments based on the observed local population of binary pulsars with probable neutron star companions lead to an estimate of the rate of neutron star binary coalescence in the Universe of order $10^{-7}\,\mathrm{yr}^{-1}\,\mathrm{Mpc}^{-3}$ (Narayan, Piran & Shemi 1991; Phinney 1991). Using this estimate, Finn & Chernoff (1993) predict that an advanced LIGO detector could observe about 70 events per year[1]. In addition to providing a major

---

[1] Theoretical models of the binary star population in our Galaxy suggest that the neutron star binary coalescence rate may be even higher, perhaps as high as

new confirmation of Einstein's theory of general relativity, the detection of gravitational waves from coalescing binaries at cosmological distances could provide the first accurate measurement of the Hubble constant and mean density of the Universe (Schutz 1986; Chernoff & Finn 1993; Marković 1993).

Recent calculations of the gravitational radiation waveforms from coalescing binaries have focused on the signal emitted during the last few thousand orbits, as the frequency sweeps upward from about 10 Hz to 1000 Hz. The waveforms in this regime can be calculated fairly accurately by performing high-order post-Newtonian expansions of the equations of motion for two *point masses*[2] (Lincoln & Will 1990; Junker & Schäfer 1992; Kidder, Will, & Wiseman 1992; Wiseman 1993; Cutler & Flanagan 1994). However, at the end of the inspiral, when the binary separation becomes comparable to the stellar radii, hydrodynamic effects become important and the character of the waveforms will change. Special purpose narrow-band detectors that can sweep up frequency in real time will be used to try to catch the corresponding final few cycles of gravitational waves (Meers 1988; Strain & Meers 1991). In this terminal phase of the coalescence, the waveforms contain information not just about the effects of general relativity, but mostly about the internal structure of the stars and the nuclear equation of state at high density. Extracting this information from observed waveforms, however, requires detailed theoretical knowledge about all relevant hydrodynamic processes.

Coalescing neutron star binaries are also at the basis of numerous models of gamma-ray bursters at cosmological distances (Paczyński 1986; Eichler et al. 1989; Narayan, Paczyński, & Piran 1992; Nemiroff 1994). The isotropic angular distribution of the bursts detected by the BATSE experiment on the Compton GRO satellite (Meegan et al. 1992) strongly suggests a cosmological origin, and the rate of gamma-ray bursts detected by BATSE, of order one per day, is in rough agreement with theoretical predictions for the rate of neutron star binary coalescence in the Universe (cf. above). The complete hydrodynamic evolution during final merging, especially in the outermost, low-density regions of the system, must be understood in detail before realistic, three-dimensional models can be constructed for the gamma-ray emission (Davies et al. 1994; Piran, these Proceedings).

---

$\gtrsim 10^{-6}$ yr$^{-1}$ Mpc$^{-3}$ (Tutukov & Yungelson 1993).

[2] High accuracy is essential here because the observed signals will be matched against theoretical templates. Since the templates must cover $\gtrsim 10^3$ orbits, a fractional error as small as $10^{-3}$ can prevent detection.

2.2. DYNAMICAL COALESCENCE

Hydrostatic equilibrium configurations for binary systems with sufficiently close components can become *dynamically unstable* (Chandrasekhar 1975; Tassoul 1975). The physical nature of this instability is common to all binary interaction potentials that are sufficiently steeper than $1/r$ (see, e.g., Goldstein 1980, §3.6). It is analogous to the familiar instability of circular orbits sufficiently close to a black hole (Shapiro & Teukolsky 1983, §12.4). Here, however, it is the *Newtonian tidal interaction* that is responsible for the steepening of the effective interaction potential between the two stars and for the destabilization of the circular orbit (LRS3).

Close binaries containing neutron stars with stiff equations of state (adiabatic exponent $\Gamma \gtrsim 2$) are particularly susceptible to this instability. This is because tidal effects are stronger for stars containing a less compressible fluid. As the dynamical stability limit is approached, the secular orbital decay driven by gravitational wave emission can be dramatically accelerated (LRS2, LRS3). The two stars then plunge rapidly toward each other, and merge together into a single object in just a few rotation periods. This dynamical instability was first identified in RS1, where we calculated the evolution of equilibrium configurations containing two identical polytropes with $\Gamma = 2$. It was found that when $r \lesssim 3R$ ($r$ is the binary separation and $R$ the radius of an unperturbed neutron star), the orbit becomes unstable to radial perturbations and the two stars undergo rapid coalescence. For $r \gtrsim 3R$, the system could be evolved dynamically for many orbital periods without showing any sign of orbital evolution (in the absence of dissipation).

The dynamical evolution of an unstable, initially synchronized (i.e., rigidly rotating) binary can be described typically as follows (RS1, RS2). During the initial, linear stage of the instability, the two stars approach each other and come into contact after about one orbital revolution. In the corotating frame of the binary, the relative velocity remains very subsonic, so that the evolution is adiabatic at this stage. This is in sharp contrast to the case of a head-on collision between two stars on a free-fall, radial orbit, where shocks are very important for the dynamics (RS1). Here the stars are constantly being held back by a (slowly receding) centrifugal barrier, and the merging, although dynamical, is much more gentle. After typically two orbital revolutions the innermost cores of the two stars have merged and the system resembles a single, very elongated ellipsoid. At this point a secondary instability occurs: *mass shedding* sets in rather abruptly. Material is ejected through the outer Lagrangian points of the effective potential and spirals out rapidly. In the final stage, the spiral arms widen and merge together. The relative radial velocities of neighboring arms as they merge are supersonic, leading to some shock-heating and dissipation. As a result, a hot, nearly axisymmetric rotating halo forms around the central dense core. No measurable amount of mass escapes from the system. The halo contains about 20% of the total mass and has a pseudo-barotropic structure (Tassoul 1978, §4.3), with the angular velocity decreasing as a power-law $\Omega \propto \varpi^{-\nu}$ where $\nu \lesssim 2$ and $\varpi$ is the distance to the rotation axis (RS1). The core is rotating uniformly near breakup speed and contains about 80% of the mass still in a cold, degenerate state.

We calculate the emission of gravitational radiation during dynamical coalescence using the quadrupole approximation (RS1). Both the frequency and amplitude of the emission peak somewhere during the final dynamical coalescence, typically just before the onset of mass shedding. Immediately after the peak, the amplitude drops abruptly as the system evolves towards a more axially symmetric state. For an initially synchronized binary containing two identical polytropes, the properties of the waves near the end of the coalescence depend very sensitively on the stiffness of the equation of state. When $\Gamma < \Gamma_{crit}$, with $\Gamma_{crit} \approx 2.3$, the final merged configuration is perfectly axisymmetric[3] and the amplitude of the waves drops to zero in just a few periods (RS1). In contrast, when $\Gamma > \Gamma_{crit}$, the dense central core of the final configuration remains *triaxial* (its structure is basically that of a compressible Jacobi ellipsoid; cf. LRS1) and therefore it continues to radiate gravitational waves. The amplitude of the waves first drops quickly to a nonzero value and then decays more slowly as gravitational waves continue to carry angular momentum away from the central core (RS2). Because realistic neutron star models give effective $\Gamma$ values precisely in the range 2—3 (LRS3), i.e., close to $\Gamma_{crit} \approx 2.3$, a simple determination of the absence or presence of persisting gravitational radiation after the coalescence (i.e., after the peak in the emission) could place a strong constraint on the stiffness of the equation of state.

### 2.3. MASS TRANSFER AND THE DEPENDENCE ON THE MASS RATIO

Clark & Eardley (1977) suggested that secular, *stable* mass transfer from one neutron star to another could last for hundreds of orbital revolutions before the lighter star is tidally disrupted. Such an episode of stable mass transfer would be accompanied by a secular *increase* of the orbital separation. Thus if stable mass transfer could indeed occur, a characteristic "reversed chirp" would be observed in the gravitational wave signal at the end of the inspiral phase (Jaranowski & Krolak 1992).

---

[3] A polytropic fluid with $\Gamma < 2.3$ (polytropic index $n > 0.8$) cannot sustain a non-axisymmetric, uniformly rotating configuration in equilibrium (see, e.g., Tassoul 1978, §10.3).

The question was reexamined recently by Kochanek (1992) and Bildsten & Cutler (1992), who both argued against the possibility of stable mass transfer on the basis that very large mass transfer rates and extreme mass ratios would be required. Moreover, in LRS3 it was pointed out that mass transfer has in fact little importance for most neutron star binaries (except perhaps those containing a very low mass neutron star). This is because for $\Gamma \gtrsim 2$, *dynamical instability always arises before the Roche limit* along a sequence of binary configurations with decreasing $r$. Therefore, by the time mass transfer begins, the system is already in a state of dynamical coalescence and it can no longer remain in a nearly circular orbit. Thus stable mass transfer from one neutron star to another appears impossible.

In RS2 we presented a complete dynamical calculation for a system containing two polytropes with $\Gamma = 3$ and a mass ratio $q = 0.85$[4]. For this system we found that the dynamical stability limit is at $r/R \approx 2.95$, whereas the Roche limit is at $r/R \approx 2.85$. The dynamical evolution turns out to be quite different from that of a system with $q = 1$. The Roche limit is quickly reached while the system is still in the linear stage of growth of the instability. Dynamical mass transfer from the less massive to the more massive star begins within the first orbital revolution. Because of the proximity of the two components, the fluid acquires very little velocity as it slides down from the inner Lagrangian point to the surface of the other star. As a result, relative velocities of fluid particles remain largely subsonic and the coalescence proceeds quasi-adiabatically, just as in the $q = 1$ case. In fact, the mass transfer appears to have essentially no effect on the dynamical evolution. After about two orbital revolutions the smaller-mass star undergoes complete tidal disruption. Most of its material is quickly spread on top of the more massive star, while a small fraction of the mass is ejected from the outermost Lagrangian point and forms a single-arm spiral outflow. The more massive star, however, remains little perturbed during the entire evolution and simply becomes the inner core of the merged configuration.

The dependence of the peak amplitude $h_{max}$ of gravitational waves on the mass ratio $q$ appears to be very strong, and nontrivial. In RS2 we obtained an approximate scaling $h_{max} \propto q^2$. This is very different from the scaling obtained for a detached binary system with a given binary separation. In particular, for two point masses in a circular orbit with separation $r$ we have $h \propto \Omega^2 \mu r^2$, where $\Omega^2 = G(M + M')/r^3$ and $\mu = MM'/(M + M')$. At constant $r$, this gives $h \propto q$. This linear scaling is obeyed (only ap-proximately, because of finite-size effects) by the wave amplitudes of the various systems at the *onset* of dynamical instability. For determining the *maximum* amplitude, however, hydrodynamics plays an essential role. In a system with $q \neq 1$, the more massive star tends to play a far less active role in the hydrodynamics and, as a result, *there is a rapid suppression of the radiation efficiency as q departs even slightly from unity*. For the peak luminosity of gravitational radiation we found approximately $L_{max} \propto q^6$. Again, this is a much steeper dependence than one would expect based on a simple point-mass estimate, which gives $L \propto q^2(1+q)$ at constant $r$.

---

[4]This is the most probable value of the mass ratio in the binary pulsar PSR 2303+46 (Thorsett et al. 1993) and represents the largest observed departure from $q = 1$ in any observed binary pulsar with likely neutron star companion. For comparison, $q = 1.386/1.442 = 0.96$ in PSR 1913+16 (Taylor & Weisberg 1989) and $q = 1.32/1.36 = 0.97$ in PSR 1534+12 (Wolszczan 1991).

## 2.4. MEASURING THE RADIUS OF A NEUTRON STAR WITH LIGO

The most important parameter that enters into quantitative estimates of the gravitational wave emission during the final coalescence is the relativistic parameter $M/R$ for a neutron star (we take $G = c = 1$). In particular, for two identical point masses we know that the wave amplitude obeys $(r_O/M)h \propto (M/R)$, where $r_O$ is the distance to the observer, and the total luminosity $L \propto (M/R)^5$. Thus one expects that any quantitative measurement of the emission near maximum should lead to a direct determination of the radius $R$, assuming that the mass $M$ has already been determined from the low-frequency inspiral waveform (Cutler & Flanagan 1994). Most current nuclear equations of state for neutron stars give $M/R \sim 0.1$, with $R \sim 10$ km nearly independent of the mass in the range $0.8 M_\odot \lesssim M \lesssim 1.5 M_\odot$ (see, e.g., Baym 1991; Cook et al. 1994; LRS3).

However, the details of the hydrodynamics also enter into this determination. The importance of hydrodynamic effects introduces an explicit dependence of all wave properties on the internal structure of the stars (which we represent here by a single dimensionless parameter $\Gamma$), and on the mass ratio $q$. If relativistic effects were taken into account for the hydrodynamics itself, an additional, nontrivial dependence on $M/R$ would also be present. This can be written conceptually as

$$\left(\frac{r_O}{M}\right) h_{max} \equiv \mathcal{H}(q, \Gamma, M/R) \times \left(\frac{M}{R}\right) \qquad (1)$$

$$\frac{L_{max}}{L_o} \equiv \mathcal{L}(q, \Gamma, M/R) \times \left(\frac{M}{R}\right)^5 \qquad (2)$$

Combining all the results of RS, we can write, in the limit where $M/R \to 0$ and for $q$ not too far from unity,

$$\mathcal{H}(q, \Gamma, M/R) \approx 2.2\, q^2 \qquad \mathcal{L}(q, \Gamma, M/R) \approx 0.5\, q^6, \qquad (3)$$

*essentially independent of* $\Gamma$ in the range $\Gamma \approx 2$–$3$ (RS2). This is in the case of synchronized spins. For nonsynchronized configurations, the spin frequency of the stars must be considered as additional parameters.

8     Frederic A. Rasio AND Stuart L. Shapiro

### 2.5. NONSYNCHRONIZED BINARIES

Recent theoretical work suggests that the synchronization time in close neutron star binaries remains always longer than the orbital decay time due to gravitational radiation (Kochanek 1992; Bildsten & Cutler 1992). In particular, Bildsten & Cutler (1992) show with simple dimensional arguments that one would need an implausibly small value of the effective viscous time, $t_{visc} \sim R/c$, in order to reach complete synchronization just before final merging. In the opposite limiting regime where viscosity is completely negligible, the fluid circulation in the binary system is conserved during the orbital decay and the stars behave approximately as Darwin-Riemann ellipsoids (Kochanek 1992; LRS3). Of particular importance are the irrotational Darwin-Riemann configurations, obtained when two initially nonspinning (or, in practice, slowly spinning) neutron stars evolve in the absence of significant viscosity. Compared to synchronized systems, these irrotational configurations exhibit smaller deviations from point-mass Keplerian behavior at small $r$. However, as shown in LRS3 and RS4, irrotational configurations for binary neutron stars with $\Gamma \gtrsim 2$ can nevertheless become dynamically unstable near contact. Thus the final coalescence of two neutron stars in a nonsynchronized binary system must still be driven by hydrodynamic instabilities.

The details of the hydrodynamics are very different, however (RS4). Because the two stars appear to be counter-spinning in the corotating frame of the binary, a vortex sheet with $\Delta v = |v_+ - v_-| \approx \Omega r$ appears when the surfaces come into contact. Such a vortex sheet is Kelvin-Helmholtz unstable on all wavelengths and the hydrodynamics is therefore rather difficult to model accurately given the limited spatial resolution of three-dimensional calculations. The breaking of the vortex sheet generates a large turbulent viscosity so that the final configuration may no longer be irrotational. In numerical simulations, however, vorticity is generated mostly through spurious shear viscosity introduced by the spatial discretization. An additional difficulty is that nonsynchronized configurations evolving rapidly by gravitational radiation emission tend to develop significant tidal lags, with the long axes of the two components becoming misaligned (LRS5). This is a purely dynamical effect, present even if the viscosity is zero, but its magnitude depends on the entire previous evolution of the system. Thus the construction of initial conditions for hydrodynamic calculations of nonsynchronized binary coalescence must incorporate the gravitational radiation reaction *self-consistently*. Instead, previous studies of nonsynchronized, equal-mass binary coalescence by Shibata, Nakamura, & Oohara (1992), Davies et al. (1994), and Zughe, Centrella, & McMillan (1994) used very approximate initial conditions consisting of two identical *spheres* (poly-tropes with $\Gamma \approx 2$) placed on an inspiral trajectory calculated for two point masses.

## 3. Coalescing White Dwarf Binaries

### 3.1. ASTROPHYSICAL MOTIVATION

Coalescing white-dwarf binaries are thought to be likely progenitors for type Ia supernovae (Iben & Tutukov 1984; Webbink 1984; Paczyński 1985; Mochkovitch & Livio 1989; Yungleson et al. 1994). To produce a supernova, the total mass of the system must be above the Chandrasekhar mass. Given evolutionary considerations, this requires two C-O or O-Ne-Mg white dwarfs. Yungelson et al. (1994) show that the expected merger rate for close pairs of white dwarfs with total mass exceeding the Chandrasekhar mass is consistent with the rate of type Ia supernovae deduced from observations. Alternatively, a massive enough merger may collapse to form a rapidly rotating neutron star (Nomoto & Iben 1985; Colgate 1990). Chen & Leonard (1993) have discussed the possibility that most millisecond pulsars in globular clusters may have formed in this way. In some cases planets may form in the disk of material ejected during the coalescence and left in orbit around the central pulsar (Podsiadlowski, Pringle, & Rees 1991). Indeed the first extrasolar planets have been discovered in orbit around a millisecond pulsar, PSR B1257+12 (Wolszczan 1994). A merger of two highly magnetized white dwarfs might lead to the formation of a neutron star with extremely high magnetic field, and this scenario has been proposed as a source of gamma-ray bursts (Usov 1992).

Close white-dwarf binaries are expected to be extremely abundant in our Galaxy. Iben & Tutukov (1984, 1986) predict that $\sim 20\%$ of all binary stars produce close pairs of white dwarfs at the end of their stellar evolution. The most common systems should be those containing two low-mass helium white dwarfs. Their final coalescence can produce an object massive enough to start helium burning. Bailyn (1993) suggests that extreme horizontal branch stars in globular clusters may be such helium-burning stars formed by the coalescence of two white dwarfs. Paczyński (1990) has proposed that the peculiar X-ray pulsar 1E 2259+586 may be the product of a recent white-dwarf merger. Planets in orbit around a massive white dwarf may also form following a merger (Livio, Pringle, & Saffer 1992).

Coalescing white-dwarf binaries are also important sources of low frequency gravitational waves that should be easily detectable by future space-based interferometers. Recent proposals for space-based interferometers include the LAGOS experiment (Stebbins et al. 1989), which should have an extremely high sensitivity (down to an amplitude $h \sim 10^{-23}$–$10^{-24}$) to sources with frequencies in the range $\sim 0.1$–$100\,\mathrm{mHz}$. Evans, Iben, &

Smarr (1987) estimate a white-dwarf merger rate of order one every 5 yr in our own Galaxy. Coalescing systems closest to Earth should produce quasi-periodic gravitational waves of amplitude $h \sim 10^{-21}$ in the frequency range $\sim 10$–$100$ mHz. In addition, the total number ($\sim 10^4$) of close white-dwarf binaries in our Galaxy emitting at lower frequencies $\sim 0.1$–$1$ mHz (the emission lasting for $\sim 10^2$–$10^4$ yr before final coalescence) should provide a continuum background signal of amplitude $h_c \sim 10^{-20}$–$10^{-21}$. Individual sources should be detectable by LAGOS above this background when their frequency becomes $\gtrsim 10$ mHz. The detection of the final burst of gravitational waves emitted during the actual merging would provide a unique opportunity to observe in "real time" the hydrodynamic interaction between the two white dwarfs, possibly followed immediately by a supernova explosion, nuclear outburst, or some other type of electromagnetic signal.

### 3.2. HYDRODYNAMICS OF COALESCING WHITE DWARF BINARIES

The results of RS3 for polytropes with $\Gamma = 5/3$ show that hydrodynamics also plays an important role in the coalescence of two white dwarfs, either because of dynamical instabilities of the equilibrium configuration, or following the onset of dynamically unstable mass transfer. Systems with $q \approx 1$ must evolve into deep contact before they become dynamically unstable and merge. Instead, equilibrium configurations for binaries with $q$ sufficiently far from unity never become dynamically unstable, but once these binaries reach their Roche limit, we find that dynamically unstable mass transfer occurs and that the less massive star is completely disrupted after a small number ($< 10$) of orbital periods (see also Benz et al. 1990). In both cases, the final merged configuration is an axisymmetric, rapidly rotating object with a core-halo structure similar to that obtained for coalescing neutron stars (RS2, RS3; see also Mochkovitch & Livio 1989).

For two massive enough white dwarfs, the merger product may be well above the Chandrasekhar mass $M_{Ch}$. The object may therefore explode as a (type Ia) supernova, or perhaps collapse to a neutron star. The rapid rotation and possibly high mass (up to $2M_{Ch}$) of the object must be taken into account for determining its final fate. Unfortunately, this is not done in current theoretical calculations of accretion induced collapse (AIC), which always consider a nonrotating white dwarf just below the Chandrasekhar limit accreting matter slowly and quasi-spherically (Canal et al. 1990; Nomoto & Kondo 1991; Isern 1994). Under these assumptions it is found that collapse to a neutron star is possible only for a narrow range of initial conditions. In most cases, a supernova explosion follows the ignition of the nuclear fuel in the degenerate core. However, the fate of a much more massive object with substantial rotational support and large deviations from spherical symmetry (as would be formed by dynamical coalescence) may be very different.

Support for this work was provided by NSF Grant AST 91–19475 and NASA Grant NAGW–2364. F. A. R. was supported by a Hubble Fellowship, funded by NASA through Grant HF-1037.01-92A from the Space Telescope Science Institute, which is operated by AURA, Inc., for NASA, under contract NAS5-26555. Computations were performed at the Cornell Theory Center, which receives major funding from the NSF and IBM, with additional support from the New York State Science and Technology Foundation and members of the Corporate Research Institute.